\documentclass[12pt, journal, draftclsnofoot, oneside, onecolumn]{IEEEtran}

\usepackage{ucs}
\usepackage[cmex10]{amsmath}
\usepackage{cite, amsfonts, amssymb, amsthm, bm, bbm, graphicx, relsize, multirow, booktabs, tikz,subfigure,soul}
\usepackage[american]{babel}
\usepackage[T1]{fontenc}
\usepackage{algorithmic, algorithm}
\usepackage[multiple]{footmisc}
\theoremstyle{definition}

\theoremstyle{remark}

\renewcommand{\IEEEQED}{\IEEEQEDopen}

\newcommand{\ds}{\displaystyle}

\begin{document}

\title{Energy Efficiency and Asymptotic Performance Evaluation of   
Beamforming Structures in  Doubly Massive MIMO mmWave Systems}
\author{Stefano Buzzi, {\em Senior Member}, {\em IEEE}, and Carmen D'Andrea
\thanks{This paper was partly presented at the {\em 2016 IEEE GLOBECOM Workshop on Emerging Technologies for 5G Wireless Cellular Networks }, Washington, DC, December 2016.}
\thanks{The authors are with the Department of Electrical and Information Engineering, University of Cassino and Lazio Meridionale, I-03043 Cassino, Italy (buzzi@unicas.it, carmen.dandrea@unicas.it).}
}
\maketitle

\begin{abstract}
Future cellular systems based on the use of millimeter waves  will heavily rely on the use of antenna arrays both at the transmitter and at the receiver.  
For complexity reasons and energy consumption issues, fully digital precoding and postcoding structures may turn out to be unfeasible, and thus suboptimal structures, making use of simplified hardware and a limited number of RF chains, have been investigated. This paper considers and makes a comparative assessment, both from a spectral efficiency and energy efficiency point of view, of several suboptimal precoding and postcoding beamforming structures for a cellular multiuser MIMO (MU-MIMO) system with large number of antennas. Analytical formulas for the asymptotic achievable spectral efficiency and for the global energy efficiency of several beamforming structures are derived in the large number of antennas regime. 
Using the most recently available data for the energy consumption of phase shifters and switches, we show that  fully-digital beamformers may actually achieve a larger energy efficiency than lower-complexity solutions, as well as that low-complexity beam-steering purely analog beamforming may in some cases represent a good performance-complexity trade-off solution. 
\end{abstract}


\section{Introduction}
The use of frequency bands at millimeter waves\footnote{Even though millimeter waves is a term that historically refers to the range 30-300 GHz, in the recent literature about future wireless networks the term is used to refer to frequency above-6 GHz, in contrapposition to the usual cellular frequencies located below 6 GHz.} (mmWaves) for cellular communications is among the most striking technological innovations brought by  fifth generation (5G)  wireless networks \cite{Buzzi5G}.  

For conventional sub-6 GHz cellular systems it has been shown that equipping a base station (BS) with a very large ($>100$) number of antennas, a technique usually referred to as \textit{Massive MIMO} \cite{Marzetta10,SILarssonMM}, brings a huge increase in the network capacity, mainly due to  the capability of serving several users on the same frequency slot with nearly orthogonal vector channels. In the massive MIMO literature, while the number of antennas at the BS grows large, the user device is usually assumed to 
have only one or very few antennas: indeed, at sub-6 GHz frequencies the wavelength is in the order of several centimeters, thus making it difficult to pack many antennas on small-sized user devices. When moving to mmWave, however, the wavelength gets reduced, and, at least in principle, a large number of antennas can be mounted not only on the BS, but also on the user device. As an example, at a carrier frequency of 30 GHz the wavelength is 1 cm, 
and for a planar antenna array with $\lambda/2$ spacing, more than 180 antennas can be placed in an area as large as a standard credit card. This leads to the concept of \textit{doubly massive MIMO system} \cite{buzziWCNC2016keynote,doubly_globecom2016}, that is defined as a wireless communication system where the number of antennas grows large at both the transmitter and  the receiver. While there are certainly a number of serious practical constraints -- e.g.,  large power consumption, low efficiency of power amplifiers, hardware complexity, ADC and beamformer implementation -- that currently prevent the feasibility of an user terminal equipped with a very large number of 
antennas, 
it is on the other hand believed that these are just technological issues that will be solved or worked around in the near future, and thus this paper presents  results on the doubly massive MIMO regime for wireless systems operating at mmWave. Doubly massive mmWave systems are envisioned to be also of interest when considering wireless fronthaul networks, and, also for vehicular communications (vehiche-to-vehicle and vehicle-to-infrastructure).

When considering MIMO architectures, and in particular massive MIMO ones, hardware complexity and energy consumption issues make the use of conventional fully digital (FD) beamforming, which requires one RF chain for each antenna element, rather prohibitive; 
as a consequence, recent research efforts have been devoted towards devising suboptimal, lower complexity, beamforming structures \cite{chih-lin-i}. In particular, hybrid (HY) beamforming structures have been proposed, with 
a limited number (much smaller than the number of antenna elements) of RF chains. The paper \cite{alkhateeb2015limited}  analyzes the achievable rate for a multiuser MIMO (MU-MIMO) system with HY pre-coding and limited feedback; it is therein shown that, for the case of single-path (i.e., rank-1) channels, HY pre-coding structures achieve a spectral efficiency very close to that of a FD beamformer. In \cite{sohrabi2016hybrid}, it is shown that a HY beamformer with a number of RF chains that is twice the number of transmitted data streams may exactly mimic a FD beamformer; the analysis, which neglects energy efficiency issues,  is  however limited to either a single-user MIMO system or a MU-MIMO system with single-antenna receivers. The paper 
\cite{SwitchesRial} proposes a new low-complexity post-coding structure, based on switches rather than on analog phase shifters; the performance of this new structure is evaluated in a rather simple scenario, i.e. single-user MIMO system with a limited number of transmit and receive antennas. 
The paper \cite{Switches_constantPS} focuses on sub-6 GHz frequencies and  introduces a novel post-coding structure made of fixed (rather than tunable) phase shifters and of switches, under the assumption that  the receiver is equipped with a large array, while the transmitters have only one antenna. 
In \cite{MIMOArchitecture}, the authors considers five different low-complexity decoding structures, all based on the use of phase shifters and switches, and provide  an analysis of the achievable spectral efficiency along with estimates of the energy consumption of the proposed structures. The paper, however, does not analyze the system energy efficiency (i.e. the ratio of the achievable rate to the consumed power \cite{buzziJSAC2016}), and focuses only on the receiver omitting a similar analysis for the transmitter implementation. 
The paper \cite{EEOptimization} considers the issue of energy efficiency in a downlink massive MIMO mmWave systems by deriving an energy-efficient HY beamformer; however, the paper considers the case in which the user terminals are equipped with just one antenna, and this is a key assumption that is exploited to solve the considered optimization problems. 
In \cite{roth2016channel}, a consumed power model for components designed for 60 GHz is given, and a comparison between FD beamforming, 1-bit ADC, and analog beamforming is given.

This paper focuses on both the achievable spectral efficiency (ASE) and the global energy efficiency (GEE) of several precoding and combining structures, ranging from the FD beamformers, to their HY analog-digital implementations, to the lower complexity purely analog (AN) beamforming structures.
The paper also proposed extensions to the considered scenario recently proposed beamforming structures such as the one based on switches \cite{SwitchesRial} and the one based on fixed phase shifters and switches \cite{Switches_constantPS}.
Putting emphasis on the large number of antennas regime, asymptotic formulas of the ASE and of the GEE are derived with reference to some of these combining structures. Special emphasis, at the analysis stage, is also given to the purely AN (beam-steering) beamformer, that can be considered as a promising solution given its extremely low complexity.   While the results on the ASE confirm, as expected, that FD beamforming achieves better performance than lower-complexity structures,  things are a little bit more involved and surprising when considering the GEE. Indeed, here, the relative ranking of the several low-complexity structures strongly depends on the adopted power consumption model for amplifiers, phase shifters, switches, etc., and our results show that, using recent power models, FD beamforming may be the most energy-efficient solution. The paper also studies the system GEE as a function of the transmitted power, and shows that there is an optimal value for the transmitted power (around the value 0 dBW in the downlink); going beyond this point increasing the transmit power is not convenient from an energy-efficient point of view since it yields a limited increase in the network throughput at the price of a faster increase of the consumed energy. 
 
This paper is organized as follows. In the next section the considered system model is described, including the clustered channel model for mmWave wireless links.
Section III contains the description of the several considered beamforming structures, while in Section IV asymptotic formulas for the system ASE and GEE of two FD beamformers,  in the limit of large number of antennas are derived, both for the uplink and downlink. Section V is entirely devoted to the exposition of  asymptotic results for the purely analog beam-steering beamformers, while in Section VI extensive numerical results are discussed. Concluding remarks are finally reported in Section VII.

\section{The system model}

The paper focuses on a single-cell MU-MIMO system wherein one BS communicates, on the same frequency slot, with several mobile users. 
The parameter $N_T$ denotes the number of transmit antennas at the BS, and $N_R$ denotes the number of receive antennas at the user's device\footnote{For the sake of simplicity all the mobile receivers are assumed to have the same number of antennas; however, this hypothesis can be easily relaxed.}. 

\subsection{The clustered channel model}

The popular narrowband clustered mmWave channel model is assumed to hold. \cite{spatiallysparse_heath,cairewsa2016,buzziTCOM2017,
lee2014exploiting}.  The baseband equivalent of the propagation channel between the transmitter and the generic receiver\footnote{For ease of notation, we omit, for the moment, the subscript $"k"$ to denote the BS to the $k$-th user channel matrix.}
is thus  represented by an $(N_R \times N_T)$-dimensional matrix expressed as:
\begin{equation}
\mathbf{H}=\gamma\sum_{i=1}^{N_{\rm cl}}\sum_{l=1}^{N_{{\rm ray},i}}\alpha_{i,l}
\sqrt{L(r_{i,l})} \mathbf{a}_r(\phi_{i,l}^r) \mathbf{a}_t^H(\phi_{i,l}^t) + \mathbf{H}_{\rm LOS}\; .
\label{eq:channel1}
\end{equation}
In Eq. \eqref{eq:channel1}, 
it is implicitly assumed that the propagation environment is made of $N_{\rm cl}$ scattering clusters, each of which contributes with $N_{{\rm ray}, i}$ propagation paths, $i=1, \ldots, N_{\rm cl}$, plus a  possibly present LOS component.  
The parameters  $\phi_{i,l}^r$ and $\phi_{i,l}^t$ denote the angles of arrival and departure of the $l^{th}$ ray in the $i^{th}$ scattering cluster, respectively. 
The quantities $\alpha_{i,l}$ and $L(r_{i,l})$ are the complex path gain and the attenuation associated  to the $(i,l)$-th propagation path. 
The complex gain  $\alpha_{i,l}\thicksim \mathcal{CN}(0, \sigma_{\alpha,i}^2)$, with  $\sigma_{\alpha,i}^2=1$  \cite{spatiallysparse_heath}. The factors $\mathbf{a}_r(\phi_{i,l}^r)$ and $\mathbf{a}_t(\phi_{i,l}^t)$ represent the normalized receive and transmit array response vectors evaluated at the corresponding angles of arrival and departure; for an uniform linear array (ULA) with half-wavelength inter-element spacing it holds
$
\mathbf{a}_t(\phi_{i,l}^t)=\displaystyle \frac{1}{\sqrt{N_T}}[1 \; e^{-j\pi \sin \phi_{i,l}^t} \; \ldots \; e^{-j\pi (N_T-1) \sin \phi_{i,l}^t}]^T$.  A similar expression can be also given for $\mathbf{a}_r(\phi_{i,l}^r)$.
Finally, $\gamma=\displaystyle\sqrt{\frac{N_R N_T}{\sum_{i=1}^{N_{\rm cl}}N_{{\rm ray},i}}}$  is a normalization factor ensuring that the received signal power scales linearly with the product $N_R N_T$.
Regarding the LOS component, denoting by 
$\phi_{\rm LOS}^r$,  $\phi_{\rm LOS}^t$,
 the arrival and departure angles corresponding to the LOS link, it is assumed that
\begin{equation}
\begin{array}{llll}
\mathbf{H}_{\rm LOS} = &  
I_{\rm LOS}(d) \sqrt{N_R N_T L(d)} e^{j \theta} \mathbf{a}_r(\phi_{\rm LOS}^r)  \mathbf{a}_t^H(\phi_{\rm LOS}^t) \; .
\end{array}
\label{eq:Hlos}
\end{equation}
In the above equation, $\theta \thicksim \mathcal{U}(0 ,2 \pi)$, while $I_{\rm LOS}(d) $ is a random variate indicating if a LOS link exists between transmitter and receiver, with $p$ the probability that $I_{\rm LOS}(d) =1$. 
A detailed description of all the parameters needed for the generation of sample realizations for the channel model of Eq. \eqref{eq:channel1} is reported in \cite{buzzidandreachannel_model}, and the reader is referred to this reference for further details on the  channel model.

\subsection{Transmitter and receiver processing}

\subsubsection{Downlink}
Assume that $M$ denotes the number of data symbols sent to each user in each signalling interval\footnote{Otherwise stated, the BS transmits in each time-frequency slot $M K$ data symbols.}, and by $\mathbf{x}_k$ the $M$-dimensional vector of the data symbols intended for the $k$-th user; the discrete-time signal transmitted by the BS can be expressed as the $N_T$-dimensional vector
$\mathbf{s}_T=  \sum_{k=1}^{K} \mathbf{Q}_k \mathbf{x}_k$, with $\mathbf{Q}_k$ the $(N_T \times M)$-dimensional pre-coding matrix for the $k$-th user. The signal received by the generic $k$-th user is expressed as the following $N_R$-dimensional vector
\begin{equation}
\mathbf{y}_k= \mathbf{H}_k \mathbf{s}_T + \mathbf{w}_k \; ,
\label{eq:yk}
\end{equation}
with $\mathbf{H}_k$ representing the clustered channel (modeled as in Eq. \eqref{eq:channel1}) from the BS to the $k$-th user and $\mathbf{w}_k$ is the $N_R$-dimensional additive white Gaussian noise with zero-mean i.i.d. entries with variance $\sigma_n^2$. Denoting by $\mathbf{D}_k$ the $(N_R \times M)$-dimensional post-coding matrix at the $k$-th user device, the following $M$-dimensional vector is finally obtained:
\begin{equation}
\mathbf{r}_k= \mathbf{D}_k^H \mathbf{H}_k \mathbf{Q}_k \mathbf{x}_k + \sum_{\ell =1, \ell \neq k}^K
\mathbf{D}_k^H \mathbf{H}_k \mathbf{Q}_\ell \mathbf{x}_\ell + \mathbf{D}_k^H \mathbf{w}_k  \; .
\label{eq:rk_downlink}
\end{equation}

\subsubsection{Uplink}
In an uplink scenario,  $K$ denotes  the number of users simultaneously transmitting to the BS in the same frequency band, and  $M$ again denotes the number of data symbols sent by each user in each signalling interval.  Letting now  $\mathbf{x}_k$ be the $M$-dimensional vector of the data symbols from the $k$-th user, the discrete-time signal transmitted by the $k$-th user device is expressed as the $N_T$-dimensional vector
$\mathbf{s}_k=  \mathbf{Q}_k \mathbf{x}_k$, with $\mathbf{Q}_k$ the $(N_T \times M)$-dimensional pre-coding matrix for the $k$-th user. The signal received by the BS is expressed as the following $N_R$-dimensional vector
\begin{equation}
\mathbf{y}= \ds \sum_{k=1}^K \mathbf{H}_k \mathbf{Q}_k \mathbf{x}_k + \mathbf{w} \; ,
\label{eq:y_uplink}
\end{equation}
with $\mathbf{H}_k$ representing now the channel from the  $k$-th user to the BS and $\mathbf{w}$  the $N_R$-dimensional additive white Gaussian noise with zero-mean i.i.d. entries with variance $\sigma_n^2$. 
Assuming, for the sake of simplicity, single-user processing at the BS, a soft estimate of the symbols from the $k$-th user is obtained as
\begin{equation}
\widehat{\mathbf{x}}_k= \mathbf{D}_k^H \mathbf{y}= \mathbf{D}_k^H \mathbf{H}_k \mathbf{Q}_k \mathbf{x}_k + \sum_{\ell =1, \ell \neq k}^K
\mathbf{D}_k^H \mathbf{H}_\ell \mathbf{Q}_\ell \mathbf{x}_\ell + \mathbf{D}_k^H \mathbf{w}  \; ,
\label{eq:xhatk}
\end{equation} 
with $\mathbf{D}_k$ the $(N_R \times M)$-dimensional post-coding matrix for the $k$-th user symbols. 
Now, depending on the choice of the pre-coding and post-coding matrices $\mathbf{Q}_k$ and $\mathbf{D}_k$,  several transceiver structures can be conceived. These will be illustrated later in the next section.

\subsection{The considered performance measures}
Two performance measures will be considered: the ASE and the GEE. 
The ASE is measured in [bit/s/Hz], while the GEE is measured in 
[bit/Joule] \cite{buzziJSAC2016}.
Assuming Gaussian data symbols\footnote{The impact on the ASE of a finite-cardinality modulation is a topic worth future investigations.} in \eqref{eq:rk_downlink}, the ASE for the downlink case can be shown to be expressed as \cite{InterferenceMIMO}
\begin{equation}
{\rm ASE}= \displaystyle \sum_{k=1}^K \log_2 \det \left[ \mathbf{I}_M + \frac{P_T}{KM}\mathbf{R}_{\overline{k}}^{-1}\mathbf{D}_k^H \mathbf{H}_k
\mathbf{Q}_k \mathbf{Q}_k^H \mathbf{H}_k^H \mathbf{D}_k
\right] \; ,
\label{eq:ASE}
\end{equation}
 wherein $\mathbf{I}_M$ is the identity matrix of order $M$, $P_T$ is the BS transmit power, and,
 according again to the signal model \eqref{eq:rk_downlink} 
 $\mathbf{R}_{\overline{k}}$ is the covariance matrix of the overall disturbance seen by the $k$-th user receiver, i.e.:
\begin{equation}
\mathbf{R}_{\overline{k}}=\sigma^2_n \mathbf{D}_k^H \mathbf{D}_k +
\frac{P_T}{MK} \sum_{\ell =1, \ell \neq k}^K
\mathbf{D}_k^H \mathbf{H}_k \mathbf{Q}_\ell \mathbf{Q}_\ell^H \mathbf{H}_k^H \mathbf{D}_k \; .
\end{equation}
For the uplink, instead, the $k$-th user ASE is expressed as \cite{InterferenceMIMO}
\begin{equation}
{\rm ASE}_k= \displaystyle  \log_2 \det \left[ \mathbf{I}_M + \frac{P_{T,k}}{M}\mathbf{R}_{\overline{k}}^{-1}\mathbf{D}_k^H \mathbf{H}_k
\mathbf{Q}_k \mathbf{Q}_k^H \mathbf{H}_k^H \mathbf{D}_k
\right] \; , \quad \forall k=1, \ldots, K \; , 
\label{eq:ASE_k}
\end{equation}
wherein $P_{T,k}$ is the $k$-th user transmit power,  
and the overall disturbance covariance matrix, according to the signal model in \eqref{eq:xhatk}, is now written as\footnote{Note that the power budget, both at the BS and at the user's transmitters, is assumed to be uniformly divided among the data streams, although power allocation could be easily performed.}
\begin{equation}
\mathbf{R}_{\overline{k}}=\sigma^2_n \mathbf{D}_k^H \mathbf{D}_k +
 \sum_{\ell =1, \ell \neq k}^K \frac{P_{T,\ell}}{M}
\mathbf{D}_k^H \mathbf{H}_\ell \mathbf{Q}_\ell \mathbf{Q}_\ell^H \mathbf{H}_\ell^H \mathbf{D}_k \; .
\end{equation}

Regarding the GEE, on the downlink it is defined as 
\begin{equation}
{\rm GEE}= \displaystyle \frac{ W {\rm ASE}}{\eta P_T + P_{\rm{TX},c}+K P_{\rm{RX},c}} \; ,
\label{eq:GEE}
\end{equation}
where  $W$ is the system bandwidth,  $P_{\rm{TX},c}$ is the amount of power consumed by the BS circuitry,  $P_{\rm{RX},c}$ is the amount of power consumed by the mobile user's device circuitry, and 
$\eta>1$ is a scalar coefficient modelling the power amplifier inefficiency. Note that, differently from what happens in the most part of existing studies on energy efficiency for cellular communications (see, for instance, references of \cite{buzziJSAC2016}), the GEE definition \eqref{eq:GEE} includes here  the power consumed both at the BS and at the mobile user's devices.  

For the uplink scenario, instead we have that the GEE of the $k$-th user is
\begin{equation}
{\rm GEE}_k= \displaystyle \frac{W{\rm ASE_k}}{\eta P_{T,k} + P_{\rm{TX},c} } \; ,
\label{eq:ASEE_k}
\end{equation}
where  $ P_{\rm{TX},c}$ is now the amount of power consumed by the $k$-th mobile device circuitry.

\section{Beamforming structures}
In the following,  the  beamforming pre-coding and post-coding structures considered in this work are detailed, along with details on their power consumption. The section mainly focuses on the downlink, although the uplink case can be treated with minor modifications.

\subsection{Channel-matched, fully-digital (CM-FD) beamforming}
Let $\mathbf{H}_k= \mathbf{U}_k \mathbf{\Lambda}_k \mathbf{V}_k^H$ denote the singular-value-decomposition (SVD) of the matrix $\mathbf{H}_k$, and assume, with no loss of generality, that the diagonal entries of $\mathbf{\Lambda}_k$ are sorted in descending order. 
The column vectors $\mathbf{u}_{k,i}$ and $\mathbf{v}_{k,i}$ denote the $i$-th column of the matrices $\mathbf{U}_k$ and $\mathbf{V}_k$, respectively.
The $k$-th user pre-coding and post-coding matrices $\mathbf{Q}_k^{\rm CM-FD}$ and  $\mathbf{D}_k^{\rm CM-FD}$  are chosen as the columns of the matrices $\mathbf{V}_k$ and 
$\mathbf{U}_k$, respectively, corresponding to the $M$ largest entries in the eigenvalue matrix $\mathbf{\Lambda}_k$, i.e.:
\begin{equation}
\mathbf{Q}_k^{\rm CM-FD}=[\mathbf{v}_{k,1} \; \mathbf{v}_{k,2} \; \ldots \; \mathbf{v}_{k,M}] \; , \qquad
\mathbf{D}_k^{\rm CM-FD}=[\mathbf{u}_{k,1} \; \mathbf{u}_{k,2} \; \ldots \; \mathbf{u}_{k,M}]\; , \quad \forall k=1, \ldots, K \; .
\end{equation}
 The CM-FD beamforming is optimal in the interference-free case, and tends to be optimal in the case in which the number of antennas at the transmitter grows large. The considered FD pre-coding architecture requires a baseband digital precoder that adapts the $M$ data streams  to the $N_T$ transmit antennas; then, for each antenna there is a  digital-to-analog-converter (DAC), an RF chain and a power amplifier (PA). At the receiver,  a low noise amplifier (LNA), an RF chain, an analog-to-digital converter (ADC) is required for each antenna, plus a baseband digital combiner that combines the $N_R$ outputs of ADC to obtain the soft estimate of the $M$ trasmitted symbols. The amount of power consumed by the transmitter circuitry can be thus expressed as:
\begin{equation}
P_{\rm{TX},c}=N_T\left(P_{\rm RFC}+P_{\rm DAC}+P_{\rm PA}\right)+P_{\rm BB} \;  ,
\end{equation} 
and the amount of power consumed by the receiver circuitry can be expressed as:
\begin{equation}
P_{\rm{RX},c}=N_R\left(P_{\rm RFC}+P_{\rm ADC}+P_{\rm LNA}\right)+P_{\rm BB} \;  .
\end{equation} 
In the above equations, $P_{\rm RFC}= 40$ mW  \cite{MIMOArchitecture} is the power consumed by the single RF chain, $P_{\rm DAC}= 110$ mW \cite{DAC_vandenbosh} is the power consumed by each DAC, 
$P_{\rm ADC}=200$mW  \cite{MIMOArchitecture} is the power consumed by each single ADC, 
$P_{\rm PA}=16 $ mW  \cite{PhasedArray60GHz} is the power consumed by the PA,
$P_{\rm LNA}=30$ mW \cite{MIMOArchitecture} is the power consumed by the LNA,  
and $P_{\rm BB}$ is the amount of power consumed by the baseband pre-coding/combiner; assuming a  CMOS implementation we have a power consumption of 243 mW \cite{BBprec_combiner}. The values of the power consumed by the single ADC present high variability in the current literature \cite{MIMOArchitecture}. A conservative value is chosen since the designs of the components reported in literature are not commercial products and as such these values might be expected to be relatively optimistic, as compared to the power consumption of the final working devices.

\subsection{Partial zero-forcing, fully digital (PZF-FD) beamforming}
Zero-forcing pre-coding nulls interference at the receiver through the constraint that the $k$-th user pre-coding be such that the product $\mathbf{H}_\ell \mathbf{Q}_k$ is zero for all $\ell \neq k$. In order to avoid a too severe noise enhancement,  a partial zero-forcing approach is adopted here, namely the columns of the pre-coding matrix $\mathbf{Q}_k$ are required to be orthogonal to the $M$ (the number of transmitted data-streams to each user) right eigenvectors of the channel $\mathbf{H}_\ell$ corresponding to the largest eigenvalues of $\mathbf{H}_\ell$, for all $\ell \neq k$. In this way, the precoder orthogonalizes only to a $M(K-1)$-dimensional subspace and nulls the most significant part of the interference.  Formally, the precoder $\mathbf{Q}_k^{\rm PZF-FD}$ is obtained as the projection of the CM-FD precoder $\mathbf{Q}_k^{\rm CM-FD}$ onto the orthogonal complement of the subspace spanned by the $M$ dominant right eigenvectors of the channel matrices $\mathbf{H}_1, \ldots, \mathbf{H}_{k-1}, \mathbf{H}_{k+1}, \ldots, \mathbf{H}_K$. The post-coding matrix is instead obtained as $\mathbf{D}_k^{\rm PZF-FD}=(\mathbf{H}_k \mathbf{Q}_k^{\rm PZF-FD})^+$, with $(\cdot)^+$ denoting pseudo-inverse. Since the PZF-FD beamforming requires a FD post-coding, its power consumption is the same as that of the CM-FD beamformer.

\subsection{Channel-matched, hybrid (CM-HY) beamforming}
\begin{figure}
  \centering
  \includegraphics[scale=0.36]{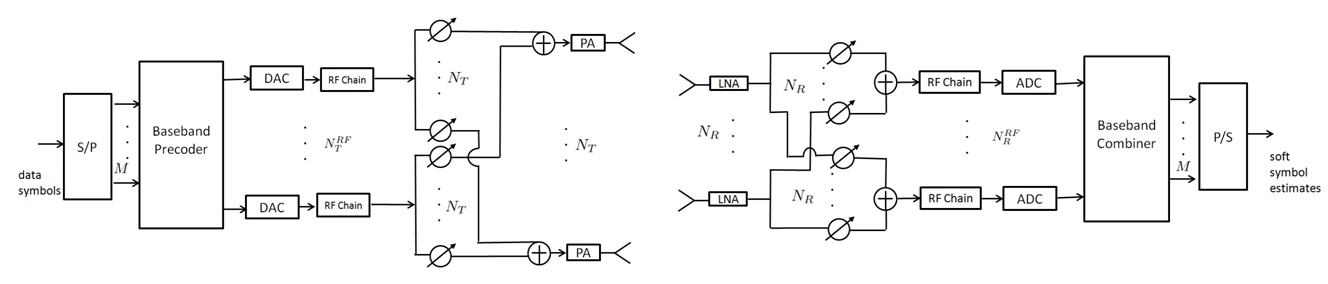}
\caption{Block-scheme of a transceiver with HY digital/analog beamforming.}
\label{Fig:hybrid_structure}
\end{figure}

In order to avoid a number of RF chains equal to the number of antennas, HY beamforming architectures have been proposed; in particular, denoting by $N_T^{\rm RF}$ and $N_R^{\rm RF}$ the number of RF chains available at the transmitter and at the receiver, respectively, the $k$-th user pre-coding and post-coding matrices are decomposed as follows:
\begin{equation}
\mathbf{Q}_k^{\rm CM-HY}=\mathbf{Q}_k^{\rm RF} \mathbf{Q}_k^{\rm BB}\; , \quad
\mathbf{D}_k^{\rm CM-HY}=\mathbf{D}_k^{\rm RF} \mathbf{D}_k^{\rm BB} \; .
\end{equation}
In the above decomposition, the matrices $\mathbf{Q}_k^{\rm RF}$ and $\mathbf{D}_k^{\rm RF}$ have dimension 
$(N_T \times N_T^{\rm RF})$ and $(N_R \times N_R^{\rm RF})$, respectively, and their entries are constrained to have constant (unit) norm (i.e. they are implemented through a network of phase-shifters\footnote{The case of quantized phase-shifts is also considered in the literature, but we are neglecting it here for the sake of simplicity.}); the matrices $\mathbf{Q}_k^{\rm BB}$ and $\mathbf{D}_k^{\rm BB}$, instead, have dimension $(N_T^{\rm RF} \times M)$ and $(N_R^{\rm RF} \times M)$, respectively, and their entries are unconstrained complex numbers. A block-scheme of the architecture of the HY transceiver is depicted in Fig. \ref{Fig:hybrid_structure}.
Now, designing an HY beamformer is tantamount to finding expressions for the matrices $\mathbf{Q}_k^{\rm RF}, \mathbf{Q}_k^{\rm BB}, \mathbf{D}_k^{\rm RF},$ and $\mathbf{D}_k^{\rm BB}$, so that some desired beamformers are approximated. For the CM-HY beamforming, the desired beamformers are the PZF-FD matrices, and their approximation is realized by using  the  block coordinate descent for subspace decomposition algorithm \cite{ghauch2015subspace}. The number of RF chains in the BS will be assumed to be equal to $KM$, while at the mobile terminal it is equal to $M$. 

The amount of power consumed by the transmitter circuitry can be now written as \cite{MIMOArchitecture}:
\begin{equation}
P_{\rm{TX},c}=N_T^{\rm RF}\left(P_{\rm RFC}+P_{\rm DAC}+N_T P_{\rm PS}\right)+N_T P_{\rm PA}+P_{\rm BB} \;  ,
\end{equation} 
and the amount of power consumed by the receiver circuitry can be expressed as:
\begin{equation}
P_{\rm{RX},c}=N_R^{\rm RF}\left(P_{\rm RFC}+P_{\rm ADC}+N_R P_{\rm PS}\right)+N_T P_{\rm LNA}+P_{\rm BB} \;  .
\end{equation}
Numerical values for the above quantities have already been given, except that for $P_{\rm PS}$, the power consumed by each  phase shifters, that is assumed to be 19.5 mW as in \cite{PhaseShifter60GHz}.

\subsection{Partial zero-forcing, hybrid (PZF-HY) beamforming}
Similarly to what has been described in the previous subsection, also the PZF beamformers may be approximated through HY architectures. In this case, expressions for the matrices $\mathbf{Q}_k^{\rm RF}, \mathbf{Q}_k^{\rm BB}, \mathbf{D}_k^{\rm RF},$ and $\mathbf{D}_k^{\rm BB}$ are to be found, so that the the PZF-FD beamforming matrices are approximated as closely as possible. Also in this case the  block coordinate descent for subspace decomposition algorithm \cite{ghauch2015subspace} can be used, and again the number of RF chains in the BS is assumed to be equal to $KM$, while at the mobile terminal it is equal to $M$. 

The amount of power consumed by the transmitter circuitry of th PZF-HY beamformers is the same as that consumed by the CM-HY ones.

\subsection{Fully Analog (AN) beam-steering beamforming}
Fully analog beamforming requires that the entries of the pre-coding and post-coding matrices have constant norm. Here, we consider an even simpler structure by introducing a further constraint and assuming that the columns of the matrices $\mathbf{Q}_k$ and $\mathbf{D}_k$ are unit-norm beam-steering vectors, i.e. the generic column of 
an $N$-dimensional beamformer is written as
\begin{equation}
\mathbf{a}(\phi)=\ds \frac{1}{\sqrt{N}}[1 \; e^{-jk d \sin\phi} \; \ldots \; e^{-jk d (N-1) \sin\phi}] \; .
\label{eq:ULA_N}
\end{equation}
Focusing on the generic $k$-th user, the columns of the matrix $\mathbf{Q}_k^{\rm AN}$ are chosen as the array responses corresponding to the departure angles in the channel model \eqref{eq:channel1} associated to the $M$ dominant paths. A similar choice is made for $\mathbf{D}_k^{\rm AN}$, whose columns contain the array responses corresponding to the $M$ arrival angles associated to the $M$ dominant paths. In order to avoid self-interference, a further constraint is added in the choice of the dominant paths to ensure that the angles of departure (arrivals) of the selected paths are spaced of at least 5 deg. 
Note that for large values of $N_T$ and $N_R$ the array responses of the transmit and receive antennas and corresponding to the departure and arrival angles associated to the dominant scatterer end up concident with dominant right and singular vectors of the channel, thus implying that the AN beamforming structure \eqref{eq:ULA_N} tends to become optimal. 
The amount of power consumed by the transmitter circuitry can be written as:
\begin{equation}
P_{\rm{TX},c}=N_T^{\rm RF}\left(P_{\rm RFC}+N_T P_{\rm element}+P_{\rm DAC}\right) \;  ,
\end{equation} 
and the amount of power consumed by the receiver circuitry can be expressed as:
\begin{equation}
P_{\rm{RX},c}=N_R^{\rm RF}\left(P_{\rm RFC}+N_R P_{\rm element}+P_{\rm ADC}\right) \;  ,
\end{equation}
where $P_{\rm element}= 27$ mW \cite{Phased_Array60GHz} is the power consumed by each element of the phased array.

\subsection{Beamforming based on switches and fixed phase shifters (SW+PHSH)}
\begin{figure}
  \centering
  \includegraphics[scale=0.35]{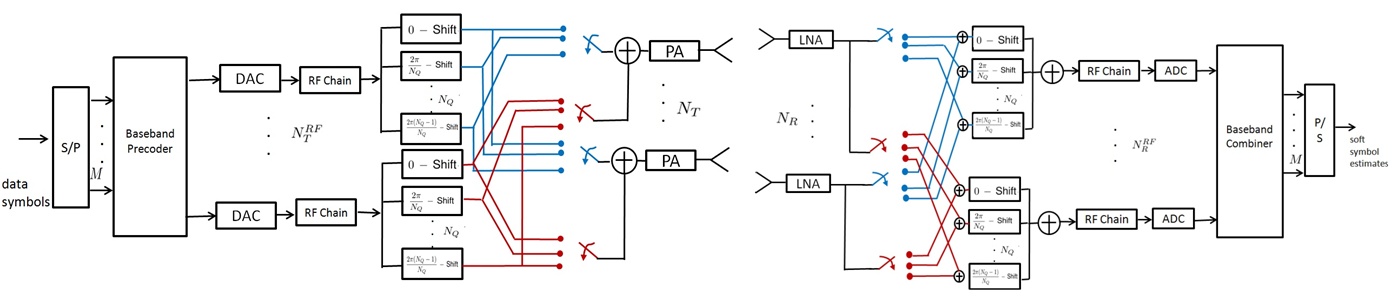}
\caption{Block-scheme of a transceiver where beamforming is implemented with switches and $N_Q$ constant phase shifters per RF chain.}
\label{Fig:switchesPS_structure}
\end{figure}

The considered structure, depicted in Fig. \ref{Fig:switchesPS_structure},  builds upon the one reported in 
\cite{Switches_constantPS}, wherein a massive MIMO combiner is proposed based on the use of switches and fixed (i.e., not tunable) phase shifters. The scheme in Fig. \ref{Fig:switchesPS_structure} extends the structure of \cite{Switches_constantPS} by including also the pre-coding design. 
The SW-PHSH beamforming structure is based on the following idea. 
The $(i,j)$-entry of the pre-coding matrix is in the form 
$[\mathbf{Q}^{\rm SW+PHSH}]_{(i,j)}=e^{j \phi_{i,j}}$, where the phase $\phi_{i,j}$ can take only discrete quantized values. It is thus an unitary module entry with a quantized phase that is obtained substituting the phase of corresponding entry of the pre-coding matrix that we aim to synthesize 
with the nearest quantizated phase, taken from the set $\left\{\frac{2(q-1)\pi}{N_Q} \, , q=1, \ldots , N_Q\right\}$.  A similar reasoning is followed for the entries of the post-coding matrix $\mathbf{D}^{\rm SW+PHSH}$. The number of quantized phases will be $N_Q=8$, and that the number of RF chains is assumed to be equal to $KM$  in the BS and to $M$ in the user' devices.   $N_Q$ constant phase shifters per RF chain are assumed, along with $N_T^{\rm RF}$ and $N_R^{\rm RF}$ switches per antenna at the transmitter and at the receiver, respectively. 

The amount of power consumed by the transmitter circuitry can be thus written as:
\begin{equation}
\begin{array}{lll}
P_{\rm{TX},c}= &N_T^{\rm RF}\left(P_{\rm RFC}+P_{\rm DAC}+N_Q P_{\rm PS}^{\rm fixed}\right)+ 
N_T\left(N_T^{\rm RF}P_{\rm SW}+P_{\rm PA}\right)+P_{\rm BB} \;  ,
\end{array}
\end{equation} 
and the amount of power consumed by the receiver circuitry can be expressed as:
\begin{equation}
\begin{array}{lll}
P_{\rm{RX},c}=&N_R^{\rm RF}\left(P_{\rm RFC}+P_{\rm ADC}+N_Q P_{\rm PS}^{\rm fixed}\right)+ 
N_R\left(N_R^{\rm RF}P_{\rm SW}+P_{\rm LNA}\right)+P_{\rm BB}  \;  .
\end{array}
\end{equation} 
In the above equations, $P_{\rm SW}=5$ mW \cite{MIMOArchitecture} is the power consumed by the single switch, and $P_{\rm PS}^{\rm fixed}$ is the power consumed by the constant phase shifter; this term  is of course lower than the power consumed by a tunable phase shifter, and is set to 1 mW.

\subsection{Switch-based (SW) beamforming}
A beamforming structure exclusively based on the use of switches is reported in  \cite{SwitchesRial}. Once again, $N_T^{\rm RF}$ and $N_R^{\rm RF}$ denote the number of RF chains at the transmitter and at the receiver, respectively, and it is assumed that there are $N_T^{\rm RF}$ switches at the transmitter and $N_R^{\rm RF}$ at the receiver that select the antennas using the Minimum Frobenius Norm (MFN) algorithm in \cite{SwitchesRial}. The pre-coding matrix is in the form $\mathbf{Q}^{\rm SW}=\mathbf{S}\mathbf{Q^{\rm BB}}$ where $\mathbf{S}$ is a $N_T \times N_T^{\rm RF}$-dimensional matrix with columns that have exactly one position containing the value "1," and the other entries in the matrix are zero, and $\mathbf{Q^{\rm BB}}$ is the $N_T^{\rm RF} \times M$-dimensional baseband pre-coding matrix. It can be thus shown that the matrix $\mathbf{Q}^{\rm SW}$ contains non-zero $N_T^{\rm RF}$ rows corresponding to the $N_T^{\rm RF}$ rows of the pre-coding matrix that we aim to synthesize with the largest norm.  A similar reasoning is followed for the entries of the post-coding matrix $\mathbf{D}^{\rm SW}$. Again, the number of RF chains in the BS is assumed to be equal to $KM$, while at the mobile terminal it is equal to $M$.

The amount of power consumed by the transmitter circuitry can be written as:
\begin{equation}
\begin{array}{lll}
P_{\rm{TX},c}=& N_T^{\rm RF}\left(P_{\rm RFC}+P_{\rm DAC}+P_{\rm SW}\right)+ 
 N_T^{\rm RF} P_{\rm PA}+P_{\rm BB} \;  ,
\end{array}
\end{equation} 
and the amount of power consumed by the receiver circuitry can be expressed as:
\begin{equation}
\begin{array}{lll}
P_{\rm{RX},c}= & N_R^{\rm RF}\left(P_{\rm RFC}+P_{\rm ADC}+P_{\rm SW}\right)+
N_R^{\rm RF} P_{\rm LNA}+P_{\rm BB}  \,  .
\end{array}
\end{equation}

\section{Asymptotic ASE and GEE analysis for the CM-FD and PZF-FD beamformers for large number of antennas}

\subsection{CM-FD beamforming, downlink}
Focusing on the downlink, 
in the large number of antennas regime, making the assumption that the set of arrival and departure angles across clusters and users are different with probability 1, 
it readily follows that  $\mathbf{D}_k^H \mathbf{H}_k \mathbf{Q}_\ell \rightarrow \mathbf{\Lambda}_{k,M}\mathbf{V}_{k,M} \mathbf{Q}_{\ell}$, whenever $k \neq \ell$,  where $\mathbf{\Lambda}_{k,M}$ is an $(M \times M)-$dimensional diagonal matrix containing the $M$ largest eigenvalues (denoted by $\lambda_{k,1}, \ldots, \lambda_{k,M}$) of the channel matrix $\mathbf{H}_k$ and 
$\mathbf{V}_{k,M}$ is an $(N_T \times M)$-dimensional matrix containing the columns of $\mathbf{V}_k$ associated to the eigenvalues in  $\mathbf{\Lambda}_{k,M}$. 
Using the above limiting values, the asymptotic ASE  in \eqref{eq:ASE} can be expressed as
\begin{equation}
{\rm ASE} \approx
\ds \sum_{k=1}^K \log_2 \det \left[ \mathbf{I}_M + \frac{P_T}{KM} 
\left(\sigma^2_n \mathbf{I}_M +
\frac{P_T}{MK} \sum_{\ell =1, \ell \neq k}^K
\mathbf{\Lambda}_{k,M} \mathbf{V}_{k,M}^H \mathbf{Q}_\ell \mathbf{Q}_\ell^H 
\mathbf{V}_{k,M} \mathbf{\Lambda}_{k,M}^H
\right)^{-1}   \mathbf{\Lambda}_{k,M} \mathbf{\Lambda}_{k,M}^H \right]
 \; .
 \label{eq:ASElimiting22}
\end{equation}
In order to explicitly show the dependence of the above formula on the number of antennas, note that the squared moduli of the  eigenvalues $\lambda_{k,i}$ depend linearly on the product $N_TN_R$. Otherwise stated, the following holds:
$
\lambda_{k,q}=\sqrt{N_TN_R} \widetilde{\lambda}_{k,q} \; , \, \forall k, q \; ,
$
with $\widetilde{\lambda}_{k,q}$ normalized eigenvalues independent of the number of transmit and receive antennas. Using this last relation, and denoting by $\{{\mu}_{k,q}\}_{q=1}^M$ the eigenvalues of the matrix
$\sum_{\ell =1, \ell \neq k}^K
\mathbf{\Lambda}_{k,M} \mathbf{V}_{k,M}^H \mathbf{Q}_\ell \mathbf{Q}_\ell^H 
\mathbf{V}_{k,M} \mathbf{\Lambda}_{k,M}^H$, straightforward manipulations lead to the following alternative expression for the ASE in \eqref{eq:ASElimiting22}:
\begin{equation}
{\rm ASE}\approx \ds \sum_{k=1}^K \sum_{q=1}^M \log_2\left[ 1+
\ds N_R N_T \ds \frac{\frac{P_T}{MK}|\widetilde{\lambda}_{k,q}|^2}{\sigma_n^2 + \frac{P_T}{MK}\mu_{k,q}}\right]\; . 
\label{eq:ASElimiting33}
\end{equation}
Eq. \eqref{eq:ASElimiting33} confirms that with the clustered channel model scenario increasing the number of antennas does not provide additional degrees of freedom but just a SINR-gain proportional to the product $N_TN_R$. 
Now, using \eqref{eq:GEE},  \eqref{eq:ASElimiting33} can be used to obtain an expression for the asymptotic GEE, i.e.:
\begin{equation}
{\rm GEE}\approx \ds \frac{\ds \sum_{k=1}^K \sum_{q=1}^M W \log_2\left[ 1+
\ds N_R N_T \ds \frac{\frac{P_T}{MK}|\widetilde{\lambda}_{k,q}|^2}{\sigma_n^2 + \frac{P_T}{MK}\mu_{k,q}}\right]}{\eta P_T + P_{\rm{TX},c}+K P_{\rm{RX},c}}
\; . 
\label{eq:GEElimiting33}
\end{equation}
An interesting problem is the GEE maximization with respect to the transmitted power, the number of transmit antennas and the number of receive antennas\footnote{Recall that $P_{\rm{TX},c}$ and $P_{\rm{RX},c}$ depend linearly on $N_T$ and $N_R$, respectively.}. While global GEE maximization with respect to $N_T, N_R$ and $P_T$ may be cumbersome, it is worth noting that a sequential iterative algorithm, wherein at each iteration maximization with respect to one parameter only is performed, can be easily conceived. Indeed, it is easily seen that the fraction in \eqref{eq:GEElimiting33} is the ratio of a concave function (with respect to the single variables $N_T, N_R$, and $P_T$) over a linear one, and, thus, Dinkelbach's algorithm may be readily applied to maximize the ratio \cite{ZapNow15}. Further details on this for the sake of brevity are not provided. However, in the section on the numerical results plots of the GEE versus $P_T$, providing an insight on the range of transmit power values that maximize the system energy efficiency, will be reported.

\subsection{CM-FD beamforming, uplink}
Similar conclusions can be also drawn for the uplink scenario. Note that in this case  $N_T$ denotes 
the number of antennas on the user's device and  $N_R$ denotes the BS array size. 
For large number of antennas, now it holds  $\mathbf{D}_k^H \mathbf{H}_\ell \mathbf{Q}_\ell \approx \mathbf{D}_k^H
\mathbf{U}_{\ell, M} \mathbf{\Lambda}_{\ell,M}$ and the $k$-th user ASE is
\begin{equation}
{\rm ASE}_k= \displaystyle  \log_2 \det \left[ \mathbf{I}_M + \frac{P_{T,k}}{M}
\left(
\sigma^2_n \mathbf{I}_M +
 \sum_{\ell =1, \ell \neq k}^K \frac{P_{T,\ell}}{M}
\mathbf{D}_k^H \mathbf{U}_{\ell,M} \mathbf{\Lambda}_{\ell,M} \mathbf{\Lambda}_{\ell,M}^H 
\mathbf{U}_{\ell,M}^H
 \mathbf{D}_k
\right)^{-1}\mathbf{\Lambda}_{k,M} \mathbf{\Lambda}_{k,M}^H \right] \; .
\label{eq:ASE_k_limiting_3}
\end{equation}
Substituting \eqref{eq:ASE_k_limiting_3} into \eqref{eq:ASEE_k} it is finally possible to obtain an asymptotic expression for the GEE of the generic $k$-th user.

\subsection{PZF-FD beamforming, downlink}

For PZF-FD beamforming, the product $\mathbf{H}_k \mathbf{Q}_\ell$ is an all-zero matrix whenever $k \neq \ell$.  As a consequence, t $\mathbf{R}_{\overline{k}} \approx \sigma_n^2 \mathbf{I}_M$, and  the asymptotic ASE can be shown to be written as\footnote{This is an asymptotic expression since we are
neglecting the noise enhancement effect (that is a decreasing function of $N_T$) induced by the nulling of the interference.}
\begin{equation}
{\rm ASE} \approx
\ds \sum_{k=1}^K \log_2 \det \left[ \mathbf{I}_M + \frac{P_T}{KM \sigma^2_n} \mathbf{\Lambda}_{k,M}\mathbf{\Lambda}_{k,M}^H \right] =
\ds \sum_{k=1}^K \sum_{q=1}^M \log_2\left[1 + \ds \frac{P_T}{MK} \frac{|\lambda_{k,q}|^2}{\sigma^2_n}\right]
\; .
\label{eq:ASElimiting}
\end{equation}
Using the normalized eigenvalues 
$
\widetilde{\lambda}_{k,q}=\lambda_{k,q}/\sqrt{N_TN_R}  \; , \, \forall k, q \; ,
$
the following equivalent expression is obtained:
\begin{equation}
{\rm ASE} \approx
\ds \sum_{k=1}^K \sum_{q=1}^M \log_2\left[1 + \ds \frac{N_T N_R P_T}{MK} \frac{|\widetilde{\lambda}_{k,q}|^2}{\sigma^2_n}\right]
\; .
\label{eq:ASElimiting2}
\end{equation}
The GEE is now written as:
\begin{equation}
{\rm GEE}\approx \ds \frac{\ds \sum_{k=1}^K \sum_{q=1}^M \log_2\left[1 + \ds \frac{N_T N_R P_T}{MK} \frac{|\widetilde{\lambda}_{k,q}|^2}{\sigma^2_n}\right]}{\eta P_T + P_{\rm{TX},c}+K P_{\rm{RX},c}}
\; ,
\label{eq:GEElimiting22}
\end{equation}
and, also in this case, Dinkelbach's algorithm can be successfully applied to perform alternative maximization of  the GEE with respect to $N_T$, $N_R$ and the transmit power $P_T$.

\subsection{PZF-FD beamforming, uplink}

Exploiting the fact that $\mathbf{D}_k^H \mathbf{H}_{\ell}$ is zero whenever $k \neq \ell$,  
the asymptotic ASE expression for the $k$-th user in this case is written as:
\begin{equation}
{\rm ASE}_k \approx
\ds \sum_{q=1}^M \log_2 \left[1 + \ds \frac{N_T N_R P_{T,k}}{M} \frac{|\widetilde{\lambda}_{k,q}|^2}{\sigma^2_n}\right]
\; ,
\label{eq:ASElimiting2_uplink}
\end{equation}
Substituting \eqref{eq:ASElimiting2_uplink} into \eqref{eq:ASEE_k} it is finally possible to obtain an asymptotic expression for the GEE of the generic $k$-th user. 

\section{Asymptotic ASE and GEE analysis for the beam-steering AN beamformers for large number of antennas}

\subsection{AN beamforming, downlink}
The case of AN pre-coding and post-coding is now considered. 
As a preliminary step to our analysis, it is convenient to recall that the ULA response in Eq. \eqref{eq:ULA_N} is a unit-norm vector, and that 
the inner product between two ULA responses of length $P$ and corresponding to incidence angles $\phi_1$ and 
$\phi_2$ is written as
\begin{equation}
f_P(\phi_1,\phi_2) \triangleq \mathbf{a}^H(\phi_1)\mathbf{a}(\phi_2) = \ds \frac{1}{P} \ds \frac{1-e^{jkd(\sin \phi_1 - \sin \phi_2)P}}{1-e^{jkd(\sin \phi_1 - \sin \phi_2)}} \; .
\end{equation}
The above inner product, that is denoted by $f_P(\phi_1,\phi_2)$, has a magnitude that, for large $P$, vanishes as $1/P$, whenever $\phi_1 \neq \phi_2$.

Let us now write the channel matrix for user $k$ as
\begin{equation}
\mathbf{H}_k=\gamma_k \ds \sum_{i=1}^N \alpha_{k,i}\mathbf{a}_r(\phi_{i,k}^r)
\mathbf{a}_t^H(\phi_{i,k}^t) = \gamma_k \mathbf{A}_{k,r}\mathbf{L}_k\mathbf{A}_{k,t}^H \; ,
\label{eq:channel_analog}
\end{equation}
namely the path-loss term has been lumped into the coefficients $\alpha_{\cdot, \cdot}$, and the summation over the clusters and the rays has been compressed in just one summation, with $N=N_{\rm cl} N_{\rm ray}$, where $N_{{\rm ray},i}=N_{\rm ray}, \; \forall i=1 \ldots N_{\rm cl}$ has been assumed.
Additionally, $\mathbf{A}_{k,r}$ is an $(N_R \times N)$-dimensional matrix containing on its columns the vectors $\mathbf{a}_r(\phi_{1,k}^r), \ldots ,  \mathbf{a}_r(\phi_{N,k}^r)$, $\mathbf{L}_k={\rm diag}(\alpha_{1,k}, \ldots, \alpha_{N,k})$, and  
$\mathbf{A}_{k,t}$ is an $(N_T \times N)$-dimensional matrix containing on its columns the vectors $\mathbf{a}_t(\phi_{1,k}^t), \ldots ,  \mathbf{a}_t(\phi_{N,k}^t)$\footnote{In order to avoid an heavy notation, we have dropped the dependence of the matrices $\mathbf{A}_{k,r}$ and $\mathbf{A}_{k,t}$ on the propagation paths arrival and departure angles, respectively.}.
It is also assumed, with no loss of generality, that the paths are sorted in decreasing magnitude order, i.e. 
$|\alpha_{1,k}| \geq |\alpha_{2,k}| \geq \ldots \geq |\alpha_{N,k}|$.  
In the following analysis, it is  assumed that there are no collisions between arrival and departure angles across users, an assumption that is usually verified unless there are very close users. 

For the downlink scenario, the analog post-coding and pre-coding matrices are written as
\begin{equation}
\mathbf{D}_k=[\mathbf{a}_r(\phi_{1,k}^r), \ldots,  \mathbf{a}_r(\phi_{M,k}^r)]  \; ,
\qquad  \mathbf{Q}_k=[\mathbf{a}_t(\phi_{1,k}^t) , \ldots,  \mathbf{a}_t(\phi_{M,k}^t)]\; ,  \quad \forall k \; ,
 \label{eq:vector_precoders_M}
 \end{equation}
and they are actually submatrices of $\mathbf{A}_{k,r}$ and $\mathbf{A}_{k,t}$, respectively.
Define now the following $(M \times N)$-dimensional matrices:
$\mathbf{F}^r_{k, \ell,M} \triangleq \mathbf{D}_k^H \mathbf{A}_{\ell,r}$ and 
$\mathbf{F}^t_{k, \ell,M} \triangleq \mathbf{Q}_k^H \mathbf{A}_{\ell,t} $
Note that the $(m,n)$-th entry of the matrix $\mathbf{F}^t_{k, \ell,M}$ is $f_{N_T}(\phi_{m,k}^t,
\phi_{n,\ell}^t)$, while the 
$(m,n)$-th entry of the matrix $\mathbf{F}^r_{k, \ell,M}$ is $f_{N_R}(\phi_{m,k}^r,
\phi_{n,\ell}^r)$. 
Equipped with this notation, the ASE in \eqref{eq:ASE} can be now expresssed as follows:
\begin{equation}
{\rm ASE}=\ds \sum_{k=1}^K 
\log_2 \det\left[
\mathbf{I}_M + \ds \frac{P_T}{KM} \gamma_k^2
\mathbf{R}_{\overline{k}}^{-1}
\mathbf{F}_{k,k,M}^r \mathbf{L}_k \mathbf{F}_{k,k,M}^{t \, H} \mathbf{F}_{k,k,M}^t 
\mathbf{L}_k^* \mathbf{F}_{k,k,M}^{r \, H} \right] \; ,
\label{eq:ASE_analog_M}
\end{equation}
with
\begin{equation}
\mathbf{R}_{\overline{k}}=
\sigma^2_n \mathbf{D}_k^H \mathbf{D}_k + \ds \frac{P_T}{MK}\gamma_k^2 \ds \sum_{\ell=1, \ell \neq k}^K
\mathbf{F}_{k,k,M}^r \mathbf{L}_k \mathbf{F}_{\ell,k,M}^{t \, H}
\mathbf{F}_{\ell,k,M}^{t} \mathbf{L}_k^* \mathbf{F}_{k,k,M}^{r \, H} \; .
\end{equation}
In order to have an asymptotic expression of Eq. \eqref{eq:ASE_analog_M} for large number of antennas, 
it can be noted that the $(M \times N)$-dimensional matrix $\mathbf{F}^r_{k,\ell,M}$ is such that (a) for $k \neq \ell$ all its entries have a norm that for large $N_R$ vanishes as $1/N_R$; while (b) for $k = \ell$ the $M$ entries on the main diagonal are equal to 1 while all the remaining terms again vanish in norm as $1/N_R$. A similar statement also applies to the matrix 
$\mathbf{F}^t_{k,\ell,M}$, of course with entries vanishing as $1/N_T$. Accordingly, the following asymptotic formulas can be proven.

\subsubsection{$N_T \rightarrow + \infty$, finite $N_R$}
In this case the system becomes interference-free and we have that
\begin{equation}
{\rm ASE}\approx \displaystyle
\sum_{k=1}^K \log_2 \det \left[\mathbf{I}_M+\ds\frac{P_T\gamma_k^2}{KM\sigma^2_n}
\left(\mathbf{D}_k^H \mathbf{D}_k\right)^{-1} \mathbf{F}_{k,k,M}^{r \, H}
\mathbf{L}_k \mathbf{L}_k^* \mathbf{F}_{k,k,M}^{r }
\right]\, .
\label{eq:ASE_limit_downlink_generalM}
\end{equation}

\subsubsection{$N_R \rightarrow + \infty$, finite $N_T$}
It holds now:
\begin{equation}
{\rm ASE}\approx\displaystyle
\sum_{k=1}^K \log_2 \det \left[\mathbf{I}_M+\ds\frac{P_T\gamma_k^2}{KM}\mathbf{R}_{\overline{k}}^{-1}\left[\mathbf{L}_k \mathbf{F}_{k,k,M}^{t \, H}
\mathbf{F}_{k,k,M}^{t} \mathbf{L}_k^*
\right]_{1:M,1:M} 
\right]\, ,
\label{eq:ASE_limit2_downlink_generalM}\end{equation}
with
\begin{equation}
\mathbf{R}_{\overline{k}}=\sigma^2_n \mathbf{I}_M+
\ds \frac{P_T\gamma_k^2}{MK}\ds \sum_{\ell=1, \ell \neq k}^K
\left[\mathbf{L}_k \mathbf{F}_{\ell,k,M}^{t \, H}
\mathbf{F}_{\ell,k,M}^{t} \mathbf{L}_k^*
\right]_{1:M,1:M}  \; .
\end{equation}

\subsubsection{$N_R, N_T \rightarrow \infty$}
Finally it holds:
\begin{equation}
{\rm ASE}\approx\displaystyle
\sum_{k=1}^K 
\sum_{\ell=1}^M
\log_2  \left[1+\ds\frac{P_T\gamma_k^2 |\alpha_{k,\ell}|^2}{KM\sigma^2_n}
\right]\, .
\label{eq:AN_downlink_limit_T_R}
\end{equation}
It is easily seen that the above expression is indeed coincident with the one reported in Eq. \eqref{eq:ASElimiting}. Additionally, substituting Eq.s \eqref{eq:ASE_limit_downlink_generalM}, 
\eqref{eq:ASE_limit2_downlink_generalM} and  \eqref{eq:AN_downlink_limit_T_R} into the GEE definition \eqref{eq:GEE}, asymptotic expressions can be readily obtain for the system GEE, and, again, these can be maximized with respect to $P_T$ by using Dinkelbach's algorithm.
\medskip

Consider now the special case $M=1$;
the pre-coding and post-coding matrices are actually column vectors, and are expressed as
\begin{equation}
\mathbf{D}_k=\mathbf{a}_r(\phi_{1,k}^r) \; , \qquad  \mathbf{Q}_k=\mathbf{a}_t(\phi_{1,k}^t) \; ,  \quad \forall k \; .
 \label{eq:vector_precoders}
 \end{equation}
Using the above expressions, it is readily seen that the ASE \eqref{eq:ASE} is written as
\begin{equation}
{\rm ASE}=\ds \sum_{k=1}^K \log_2 \left[
1+ \ds \frac{P_T}{M}\mathbf{R}_{\overline{k}}^{-1} \left|\mathbf{a}_r^H(\phi_{1,k}^r) \mathbf{H}_k 
\mathbf{a}_t(\phi_{1,k}^t) \right|^2 \right] \; .
\label{eq:ASE_downlink_analog}
\end{equation}
The interference covariance matrix $\mathbf{R}_{\overline{k}}$ is now just a scalar, and is written as
\begin{equation}
\mathbf{R}_{\overline{k}}=\sigma^2_n + \ds \sum_{\ell=1, \ell \neq k}^K \frac{P_T}{KM} \gamma_k^2
\left|\alpha_{k,1} f_{N_T}(\phi_{1,k}^t, \phi_{1, \ell}^t)   + 
\ds \sum_{i=2}^N \alpha_{k,i} f_{N_R}(\phi_{1,k}^r, \phi_{i,k}^r) 
f_{N_T}(\phi_{i,k}^t, \phi_{1,\ell}^t)
\right|^2 \; .
\label{eq:cov_int_downlink_analog}
\end{equation}
Substituting Eq. \eqref{eq:cov_int_downlink_analog} into Eq. \eqref{eq:ASE_downlink_analog}, and elaborating, the following expression is obtained:
\begin{equation}
{\rm ASE}=\ds \sum_{k=1}^K
\log_2 \left[ 1 + \ds  \ds \frac
{\ds
\frac{P_T}{KM}\gamma_k^2 \left|
\alpha_{k,1} + \ds \sum_{i=2}^N \alpha_{k,i} f_{N_R}(\phi_{1,k}^r, \phi_{i,k}^r) f_{N_T}
( \phi_{i,k}^t, \phi_{1 k}^t
)
\right|^2
}
{\sigma^2_n + \ds \sum_{\ell=1, \ell \neq k}^K \frac{P_T}{KM} \gamma_k^2
\left|\alpha_{k,1} f_{N_T}(\phi_{1,k}^t, \phi_{1, \ell}^t)   + 
\ds \sum_{i=2}^N \alpha_{k,i} f_{N_R}(\phi_{1,k}^r, \phi_{i,k}^r) 
f_{N_T}(\phi_{i,k}^t, \phi_{1,\ell}^t)
\right|^2}\right]
\label{eq:ASE_downlink_analog2}
\end{equation}
Equation \eqref{eq:ASE_downlink_analog2} provides the exact downlink  ASE expression for finite values of $N_T$ and 
$N_R$ in the case of analog pre-coding and decoding, as a function of the reflection coefficients $\alpha_{\cdot, \cdot}$ and of the departure and arrival angles. In order to study its asymptotic values for large $N_R$ and $N_T$, recall that $\gamma_k^2=N_RN_T/N$.

\subsubsection{$N_T \rightarrow + \infty$, finite $N_R$}
In this case the following holds
\begin{equation}
{\rm ASE}\approx \ds \sum_{k=1}^K\log_2 \left[ 1 + \ds \frac{P_T}{KM}\frac{|\alpha_{k,1}|^2 N_T N_R}{N \sigma^2_n}\right] \; .
\label{eq:ASE_limit_hybrid_downlink1}
\end{equation}
It is seen that the ASE grows linearly with the number of users, logarithmically with the product $N_TN_R$ , and the system is asymptotically interference-free and noise-limited. It can be also verified that the limiting ASE in Eq. \eqref{eq:ASE_limit_hybrid_downlink1} tends to coincide with the limiting ASE reported in Eq. \eqref{eq:ASElimiting2}, which holds for the case of FD beamforming, thus confirming the optimality of the considered analog beamforming in the limit of large number of transmit antennas.

\subsubsection{$N_R \rightarrow + \infty$, finite $N_T$}
In this case the following holds
\begin{equation}
\begin{array}{lll}
{\rm ASE}
& \approx  \ds \sum_{k=1}^K
\log_2 \left[ 1 + \ds  \ds \frac
{\ds
\frac{P_T}{KM}\frac{N_RN_T}{N} \left|
\alpha_{k,1} 
\right|^2
}
{\sigma^2_n + \ds \sum_{\ell=1, \ell \neq k}^K \frac{P_T}{KM} \frac{N_RN_T}{N}
\left| f_{N_T}(\phi_{1,k}^t, \phi_{1, \ell}^t) \right|^2 \left| \alpha_{k,1} 
\right|^2}\right]  
\\ & 
\rightarrow
\ds \sum_{k=1}^K
\log_2 \left[ 1 + \ds  \ds \frac
{1
}
{ \ds \sum_{\ell=1, \ell \neq k}^K 
\left| f_{N_T}(\phi_{1,k}^t, \phi_{1, \ell}^t) \right|^2} \right] 
\; .
\end{array}
\label{eq:ASE_limit_hybrid_downlink2}
\end{equation}
The ASE  converges towards an asymptote that is independent of the number of receive antennas, while the system is now noise-free and interference-limited.  The ASE now increases logarithmically with $N_T^2$, and there is no longer a linear increase of the ASE in the number of users. 
In particular, since, for large $K$ we have that the quantity $\ds \sum_{\ell=1, \ell \neq k}^K 
\left| f_{N_T}(\phi_{1,k}^t, \phi_{1, \ell}^t) \right|^2$ converges to $(K-1) E\left[ \left| f_{N_T}(\phi_{1,k}^t, \phi_{1, \ell}^t) \right|^2
\right]$, it can be shown that
\begin{equation}
\ds 
\lim_{N_R, K \rightarrow +\infty} {\rm ASE}=  \ds \frac{(\ln 2)^{-1}}{ E\left[ \left| f_{N_T}(\phi_{1,k}^t, \phi_{1, \ell}^t) \right|^2
\right]} \; .
\end{equation}
Note that the above limiting value increases with $N_T^2$, while, for large $K$,  the ASE per user vanishes.

\subsubsection{$N_R, N_T \rightarrow \infty$}
In this case the same results as in \textit{4)} hold.

\subsection{AN beamforming, uplink}
For the uplink scenario, using the notation previously introduced, the $k$-th user ASE can be shown to be expressed as
\begin{equation}
{\rm ASE}_k= 
\log_2 \det\left[
\mathbf{I}_M + \ds \frac{P_{T,k}}{M} \gamma_k^2
\mathbf{R}_{\overline{k}}^{-1}
\mathbf{F}_{k,k,M}^r \mathbf{L}_k \mathbf{F}_{k,k,M}^{t \, H} \mathbf{F}_{k,k,M}^t 
\mathbf{L}_k^* \mathbf{F}_{k,k,M}^{r \, H} \right] \; ,
\label{eq:ASEk_analog_M}
\end{equation}
with
\begin{equation}
\mathbf{R}_{\overline{k}}=
\sigma^2_n \mathbf{D}_k^H \mathbf{D}_k + \ds  \ds \sum_{\ell=1, \ell \neq k}^K
\frac{P_{T,\ell}}{M}\gamma_\ell^2
\mathbf{F}_{k,\ell,M}^r \mathbf{L}_\ell \mathbf{F}_{\ell,\ell,M}^{t \, H}
\mathbf{F}_{\ell,\ell,M}^{t} \mathbf{L}_\ell^* \mathbf{F}_{k,\ell,M}^{r \, H} \; .
\end{equation}
Asymptotic approximations for the $k$-th user ASE are now provided.

\subsubsection{$N_R \rightarrow + \infty$, finite $N_T$}
In this case the system becomes interference-free and the following holds:
\begin{equation}
{\rm ASE}_k\approx \displaystyle
\log_2 \det \left[\mathbf{I}_M+\ds\frac{P_{T,k}\gamma_k^2}{M\sigma^2_n}
\left[ 
\mathbf{L}_k \mathbf{F}_{k,k,M}^{t \, H} \mathbf{F}_{k,k,M}^{t} \mathbf{L}_k^* \right]_{1:M,1:M}
\right]\, .
\label{eq:ASE_limit_UL_NR}
\end{equation}

\subsubsection{$N_T \rightarrow + \infty$, finite $N_R$}
We have now:
\begin{equation}
{\rm ASE}_k\approx\displaystyle
\log_2 \det \left[\mathbf{I}_M+\ds\frac{P_{T,k}\gamma_k^2}{M}\mathbf{R}_{\overline{k}}^{-1}
\mathbf{F}_{k,k,M}^r \mathbf{L}_k \mathbf{L}_k^*
\mathbf{F}_{k,k,M}^{r \, H}
\right]\, ,
\label{eq:ASE_limit_UL_NT}
\end{equation}
with
\begin{equation}
\mathbf{R}_{\overline{k}}=\sigma^2_n (\mathbf{D}_k^H \mathbf{D}_k)+
\ds \sum_{\ell=1, \ell \neq k}^K
\ds \frac{P_{T,\ell}\gamma_\ell^2}{M}
\mathbf{F}_{k,\ell,M}^r \mathbf{L}_{\ell} \mathbf{L}_{\ell}^*
\mathbf{F}_{k,\ell,M}^{r\, H}
  \; .
\end{equation}

\subsubsection{$N_R, N_T \rightarrow \infty$}
Finally, the following holds:
\begin{equation}
{\rm ASE}_k\approx \displaystyle
\sum_{\ell=1}^M
\log_2  \left[1+\ds\frac{P_{T,k}\gamma_k^2 |\alpha_{k,\ell}|^2}{M\sigma^2_n}
\right]\, .
\end{equation}
The above equation can be seen to be equal to Eq. \eqref{eq:ASElimiting2_uplink}.

\medskip

Similarly to the downlink, also for the uplink the case $M=1$, considered below,  permits skipping the matrix notation and obtaining more insightful formulas. 
For $M=1$ the pre-coding and post-coding vectors are still given by Eq. \eqref{eq:vector_precoders}, and the $k$-th user ASE in Eq. \eqref{eq:ASE_k}, after some algebra, 
 is written as
\begin{equation}
{\rm ASE}_k=\ds 
\log_2 \left[ 1 + \ds  \ds \frac
{\ds
\frac{P_{T,k}}{M}\gamma_k^2 \left|
\alpha_{k,1} + \ds \sum_{i=2}^N \alpha_{k,i} f_{N_R}(\phi_{1,k}^r, \phi_{i,k}^r) f_{N_T}
( \phi_{i,k}^t, \phi_{1 k}^t
)
\right|^2
}
{\sigma^2_n + 
\ds \sum_{\ell=1, \ell \neq k}^K \frac{P_{T,\ell}}{M} \gamma_\ell^2
\left| \alpha_{\ell,1}f_{N_R}(\phi_{1,k}^r, \phi_{1, \ell}^r)  + 
\ds \sum_{i=2}^N \alpha_{\ell,i} f_{N_T}(\phi_{i,\ell}^t, \phi_{1,\ell}^t) 
f_{N_R}(\phi_{1,k}^r, \phi_{i,\ell}^r)
\right|^2}\right]
\label{eq:ASE_downlink_analog2_uplink}
\end{equation}
Equation \eqref{eq:ASE_downlink_analog2} provides the downlink  ASE expression for finite values of $N_T$ and 
$N_R$ in the case of analog pre-coding and decoding, as a function of the reflection coefficients 
$\alpha_{\cdot, \cdot}$ and of the departure and arrival angles. In order to study its asymptotic values for large $N_R$ and $N_T$, recall that $\gamma_k^2=N_RN_T/N$.

\subsubsection{$N_R \rightarrow + \infty$, finite $N_T$}
In this case the following expression holds:
\begin{equation}
{\rm ASE}_k \rightarrow \ds \log_2\left[ 1 + \ds \frac{P_{T,k}}{M}\frac{|\alpha_{k,1}|^2 N_T N_R}{N \sigma^2_n}\right] \; .
\label{eq:ASE_limit_hybrid_uplink1}
\end{equation}
It is seen that the ASE grows linearly with the number of users, logarithmically with the product $N_TN_R$ , and the system is asymptotically interference-free and noise-limited. It can be also verified that the limiting ASE in Eq. \eqref{eq:ASE_limit_hybrid_uplink1} tends to coincide with the limiting ASE reported in Eq. \eqref{eq:ASElimiting2}, which holds for the case of FD beamforming, thus confirming the optimality of the considered analog beamforming in the limit of large number of transmit antennas.

\subsubsection{$N_T \rightarrow + \infty$, finite $N_R$}
In this case the following relation holds:
\begin{equation}
\begin{array}{lll}
{\rm ASE}_k
& \approx  
\log_2 \left[ 1 + \ds  \ds \frac
{\ds
\frac{P_{T,k}}{M}\frac{N_RN_T}{N} \left|
\alpha_{k,1} 
\right|^2
}
{\sigma^2_n + \ds \sum_{\ell=1, \ell \neq k}^K \frac{P_{T, \ell}}{M} \frac{N_RN_T}{N}
\left| f_{N_R}(\phi_{1,k}^r, \phi_{1, \ell}^r) \right|^2 \left| \alpha_{\ell,1} 
\right|^2}\right] 
 \\ & 
 \rightarrow
\log_2 \left[ 1 + \ds  \ds \frac
{ \left|
\alpha_{k,1} 
\right|^2
}
{ \ds \sum_{\ell=1, \ell \neq k}^K 
\left|
\alpha_{\ell,1} 
\right|^2\left| f_{N_R}(\phi_{1,k}^r, \phi_{1, \ell}^r) \right|^2} \right] 
\; .
\end{array}
\label{eq:ASE_limit_hybrid_uplink2}
\end{equation}
The ASE  converges towards an asymptote that is independent of the number of transmit antennas, while the system is now noise-free and interference-limited.  The ASE now increases logarithmically with $N_R^2$.
For large $K$, the following holds:
$$
\ds \sum_{\ell=1, \ell \neq k}^K 
\left|
\alpha_{\ell,1} 
\right|^2\left| f_{N_R}(\phi_{1,k}^r, \phi_{1, \ell}^r) \right|^2 \approx 
(K-1) E\left[
\left|
\alpha_{\ell,1} 
\right|^2\left| f_{N_R}(\phi_{1,k}^r, \phi_{1, \ell}^r) \right|^2
\right] \; ,
$$
and also in this case the ASE per user vanishes. 

\subsubsection{$N_R, N_T \rightarrow \infty$}
In this case the same results as in \textit{4)} hold.

\section{Numerical results}
Simulation results showing the ASE and the GEE for a single-cell mmWave MIMO system are now provided; it is assumed that there are $K=10$ users using the same frequency band and whose locations  are random, with $100$ m maximum distance from the BS. The parameters for the generation of the matrix channels are the ones reported in \cite{buzzidandreachannel_model}
for the ``street canyon model'', with $N_{\rm cl}=2$ and $N_{\rm ray}=20$.
The  carrier frequency is $f_c=73$ GHz, the used bandwidth is $W=500$ 
MHz\footnote{Standardization bodies have not yet set the mmWave carrier frequencies that will be really used in practice. However, the considered values can be deemed as representative of a typical mmWave link for wireless cellular communications.},  
the noise power $\sigma^2_n= F {\cal N}_0 W$, with $F=3$ dB the receiver noise figure and 
${\cal N}_0= -174$ dBm/Hz. All the considered low-complexity beamformers have been realized using a number of RF chains equal to the multiplexing order $M$. The shown results come from an average over 500 independent realizations of users' locations and propagation channels.

\begin{figure}[t]
\centering
\includegraphics[scale=0.80]{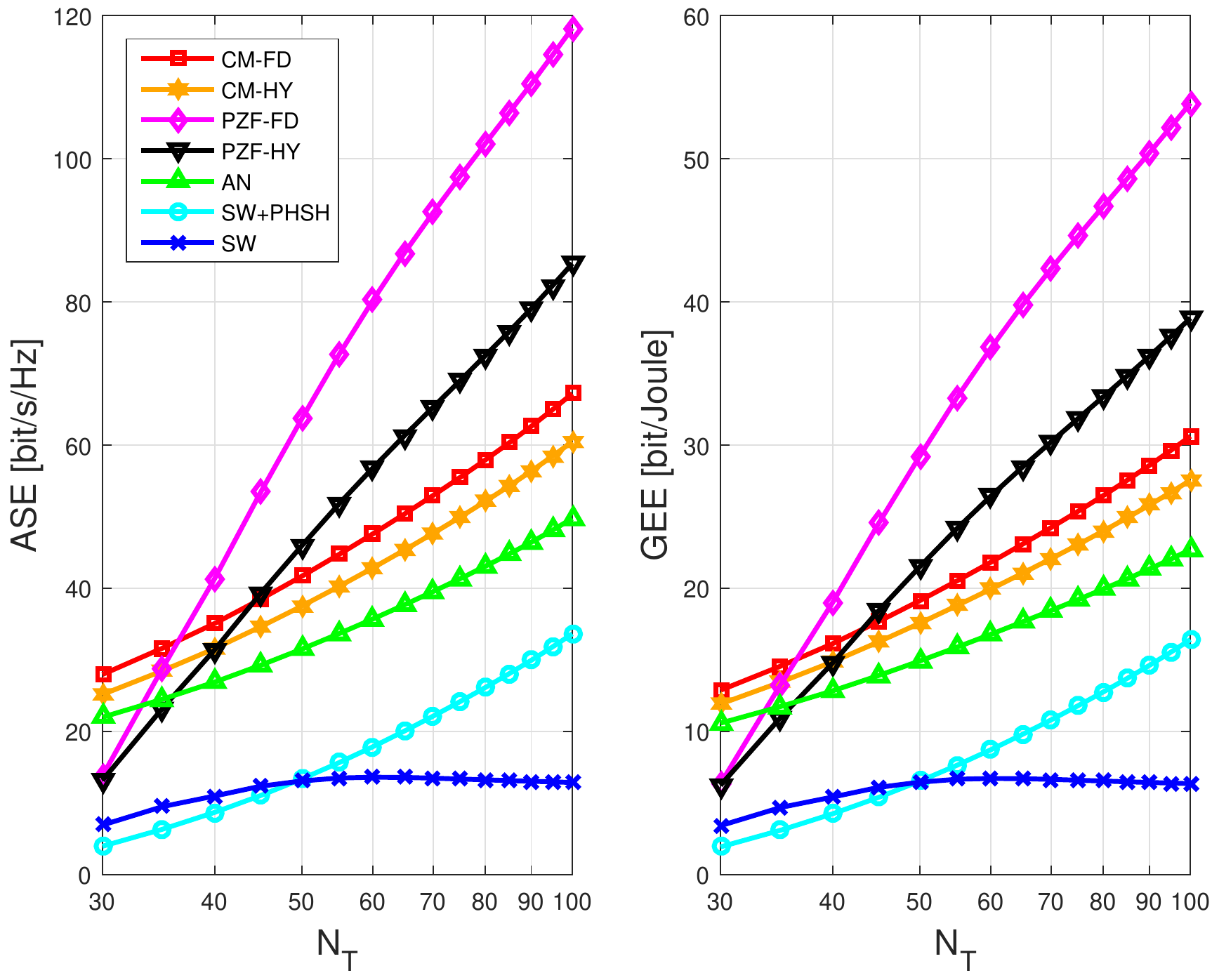}
\caption{Plot of downlink ASE and GEE versus $N_T$ with $N_R=30$, $K=10$, $M=3$ and $P_T=0$ dBW.}
\label{Fig:fig_NT}
\end{figure}

\begin{figure}[t]
\centering
\includegraphics[scale=0.80]{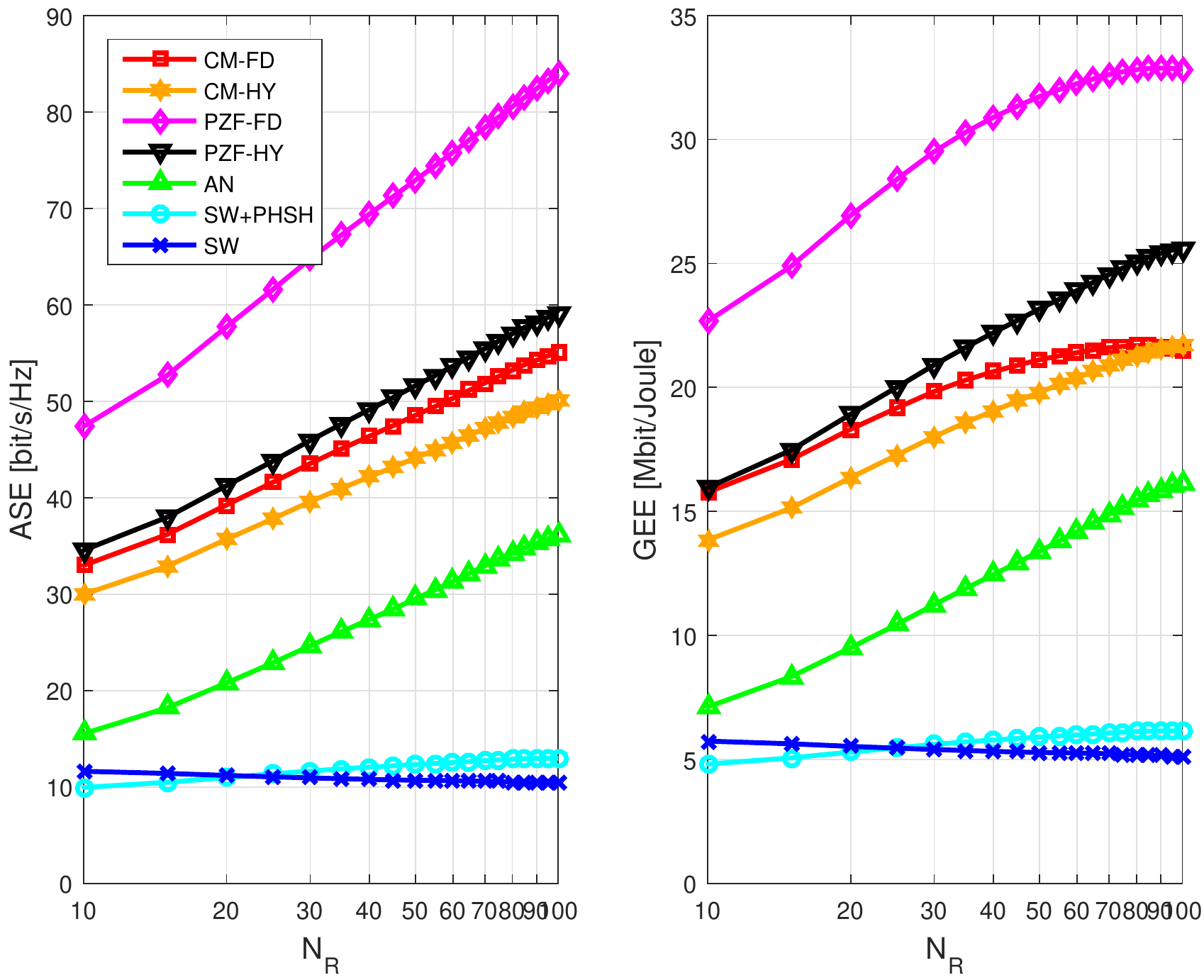}
\caption{Plot of downlink ASE and GEE versus $N_R$ with $N_T=50$, $K=10$, $M=3$ and $P_T=0$ dBW.}
\label{Fig:fig_NR}
\end{figure}

First of all, results as a function of the number of transmit and receive antennas are reported. 
Figs. \ref{Fig:fig_NT} and \ref{Fig:fig_NR} report the downlink ASE and the GEE versus the number of transmit antennas (assuming $N_R=30$) and versus the number of receive antennas (assuming $N_T=50$), respectively, assuming $P_T=0$ dBW and multiplexing order  $M=3$. Results corresponding to all the previously detailed beamforming structures are reported.  
 Inspecting the figures, it is seen that the best performing beamforming structure is the PZF-FD, both in terms of ASE and of GEE\footnote{Note however that for small values of $N_T$ the PZF beamforming structures
 achieve inferior performance with respect to the other solutions due to the reduced dimensionality of the interference-free subspace.}. This last conclusion is quite surprising, since it shows that lower complexity structures, although necessary for obvious practical considerations, actually are less energy efficient (from a communication physical layer perspective) than FD structures. Results also show that the SW structure achieves quite unsatisfactory performance; moreover,  for low values of $N_T$  the CM-FD and its HY approximation outperform the PZF-FD and  PZF-HY solutions. From Fig. \ref{Fig:fig_NR} it can be also seen that while the ASE grows with the number of antennas, the GEE instead exhibits a maximum: in particular, it is seen that, for the considered scenario, the PZF-FD beamformer achieves its maximum GEE for $N_T \approx 90$. 
 
\begin{figure}[t]
\centering
\includegraphics[scale=0.80]{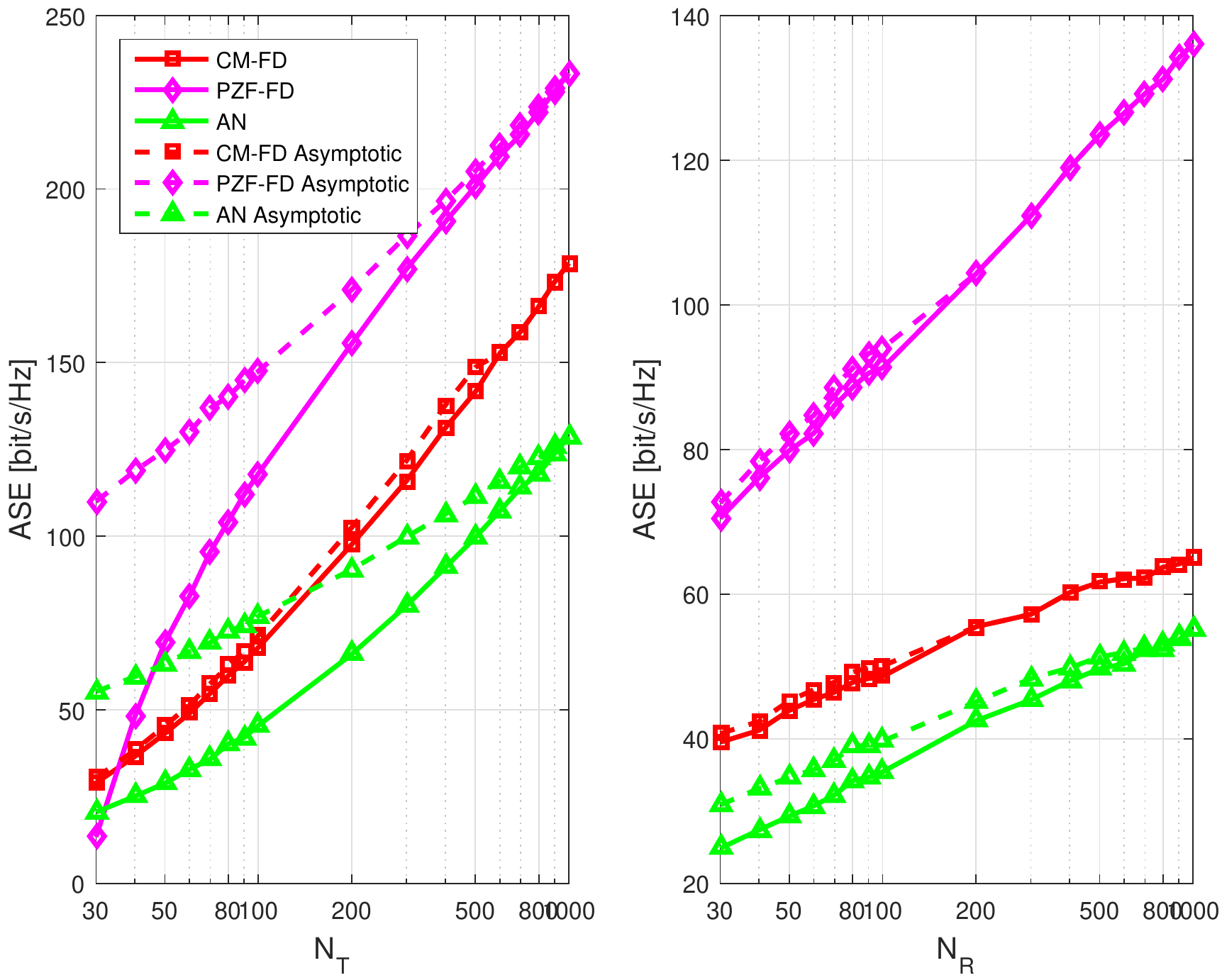}
\caption{Plot of asymptotic downlink ASE formulas versus $N_T$ and $N_R$, with $K=10$, $M=3$ and $P_T=0$ dBW.}
\label{Fig:asympDL}
\end{figure}

\begin{figure}[t]
\centering
\includegraphics[scale=0.80]{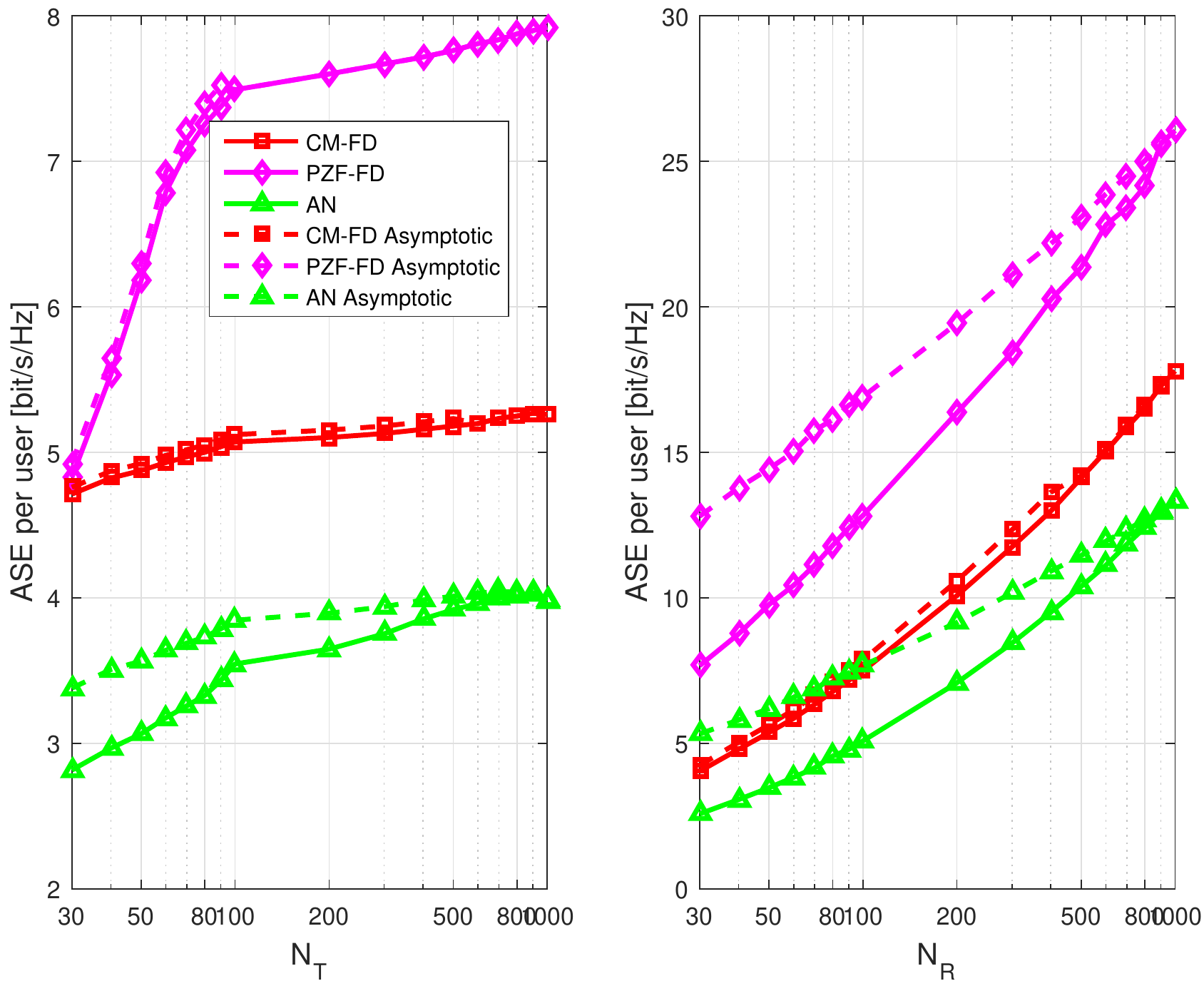}
\caption{Plot of asymptotic uplink ASE per user formulas versus $N_T$ and $N_R$, with $K=10$, $M=3$ and $P_T=0$ dBW.}
\label{Fig:asympUL}
\end{figure}

 Figs. \ref{Fig:asympDL} and \ref{Fig:asympUL} are devoted to the validation of the derived asymptotic formulas in the large number of antennas regime. In particular, the subplots in \ref{Fig:asympDL} show the downlink ASE, versus $N_T$ (assuming $N_R=30)$ and versus $N_R$ (assuming $N_T=50$), for the CM-FD, PZF-FD and AN beamformers, and their asymptotic approximation reported in  \eqref{eq:ASElimiting33},
\eqref{eq:ASElimiting2},
\eqref{eq:ASE_limit_downlink_generalM}, and 
\eqref{eq:ASE_limit2_downlink_generalM}.  The subplots in Fig. \ref{Fig:asympUL}, instead, refer to the uplink and report the ASE per user, again versus $N_T$ (assuming $N_R=30)$ and versus $N_R$ (assuming $N_T=50$), for the CM-FD, PZF-FD and AN beamformers, and their asymptotic approximation reported in 
\eqref{eq:ASE_k_limiting_3},
\eqref{eq:ASElimiting2_uplink},
\eqref{eq:ASE_limit_UL_NR}, and 
\eqref{eq:ASE_limit_UL_NT}. Results fully confirm the effectiveness of the found asymptotic formulas, that may turn out to be useful in the derivation of simplifies resource allocation strategies. 

\begin{figure}[t]
\centering
\includegraphics[scale=0.80]{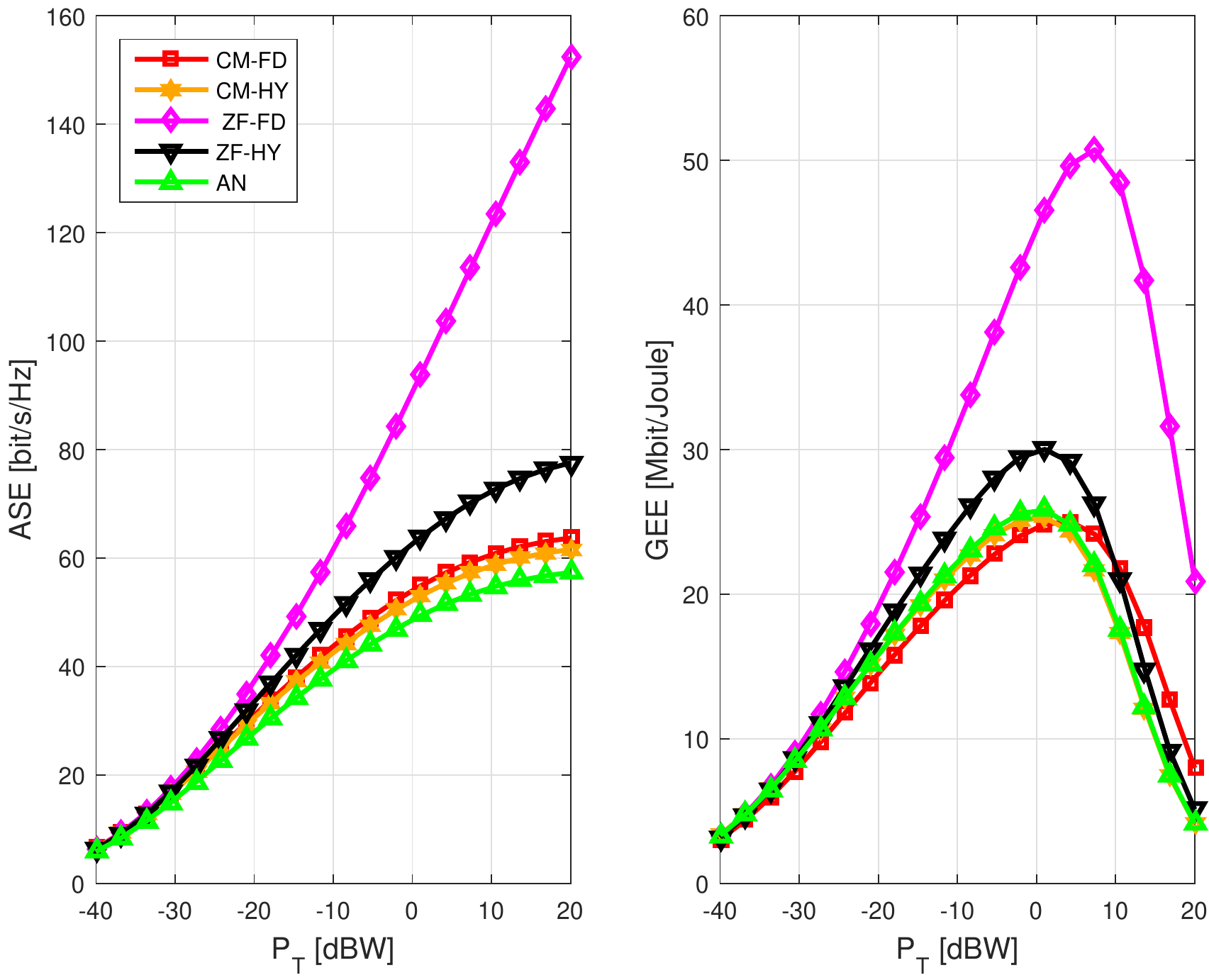}
\caption{Plot of downlink ASE and GEE versus $P_T$ for a system with $N_T=100$, $N_R=30$, $K=10$ and $M=1$.}
\label{Fig:ASEGEEM1}
\end{figure}

\begin{figure}[t]
\centering
\includegraphics[scale=0.80]{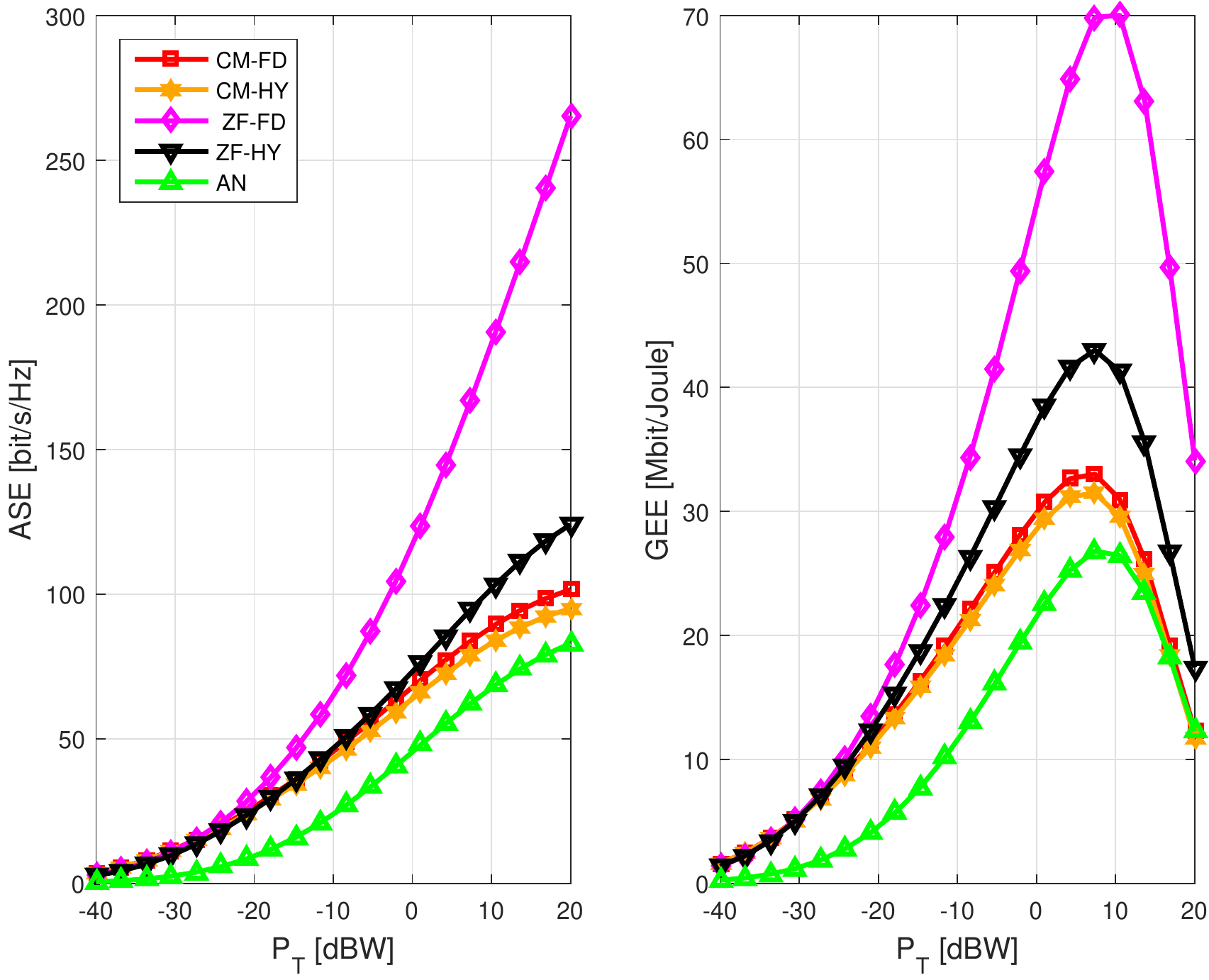}
\caption{Plot of downlink ASE and GEE versus $P_T$ for a system with $N_T=100$, $N_R=30$, $K=10$ and $M=3$.}
\label{Fig:ASEGEEM3}
\end{figure}

Figs. \ref{Fig:ASEGEEM1} and \ref{Fig:ASEGEEM3}, finally, report the downlink system ASE and GEE versus the transmit power, for the case of multiplexing order $M=1$ and $M=3$. Here, a system with $N_T=100$ antennas at the BS and $N_R=30$ antennas at the MSs has been considered. Also in this case the number of users is $K=10$. Results show here a trend that has already been found elsewhere (e.g., in \cite{ZapTWC14}); in particular, while the ASE grows with the transmit power (at least in the considered range of values), the GEE exhibits instead a maximum around 0 dBW. This behavior is explained by the fact that for large values of the transmit power, the numerator in the GEE grows at a slower rate than the denominator of the GEE, and so the GEE itself decreases. From an energy-efficient perspective, increasing the transmit power beyond the GEE-optimal point  leads to moderate improvements in the system throughput at the price of a much higher increase in the consumed power.

\section{Concluding remarks}
The paper has provided an analysis of both the spectral efficiency and the energy efficiency of a multiuser doubly massive MIMO system operating at mmWave frequencies and with several FD and low-complexity beamforming architectures. The obtained results have revealed that, using some of the most recent available data on the energy consumption of transceiver components, FD architectures, although unfeasible for large number of antennas due to complexity constraints, are superior not only in terms of achievable rate (as it was largely expected), but also in terms of energy efficiency. In particular, among FD implementations, the PZF-FD architecture has been shown to provide the best performance, while, among the lower complexity implementations, the  AN structure can be considered for its extremely low complexity. 
A detailed analytical study of some beamforming structures in the large number of antennas regime has also been provided, and results have been shown proving the accuracy of the found approximations. 
Of course the provided results and the relative ranking among the considered structures  in terms of energy efficiency is likely to change in the future as technology progresses and devices with reduced power consumption appear on the scene, even though it may be expected that in the long run FD architectures will be fully competitive, in terms of hardware complexity and energy consumption, with HY alternatives.
The analysis provided in this paper has assumed a uniform power splitting among users and data-streams; it is expected that improved performance can be obtained through waterfilling-like power control. Moreover, Gaussian-distributed data symbols have been assumed, while, instead, the effect of finite-cardinality modulation is also worth being investigated. Finally, the provided results have been derived under the assumption of perfect channel state information, and it is thus of interest to extend this work to the case in which there are channel estimation errors.

\bibliographystyle{IEEEtran}

\linespread{1.33}
\bibliography{FracProg_SB,finalRefs,references}

\end{document}